\documentclass[12pt,aps,showpacs,preprint,preprintnumbers,amsmath,amssymb,nofootinbib]{revtex4}
\usepackage{epsfig}
\usepackage{graphicx}
\usepackage{dcolumn}
\usepackage{bm}

\def\beq{\begin{equation}}
\def\eeq{\end{equation}}

\newcommand{\bea}{\begin{eqnarray}}
\newcommand{\eea}{\end{eqnarray}}

\let\jnfont=\rm
\def\NPB#1 {{\jnfont Nucl.\ Phys.\ B\ }{\bf #1}}
\def\PLB#1 {{\jnfont Phys.\ Lett.\ B\ }{\bf #1\ }}
\def\EPJC#1 {{\jnfont Eur.\ Phys.\ Jour.\ C\ }{\bf #1}}
\def\PRD#1 {{\jnfont Phys.\ Rev.\ D\ }{\bf #1\ }}
\def\PRL#1 {{\jnfont Phys.\ Rev.\ Lett.\ }{\bf #1\ }}
\def\MPLA#1 {{\jnfont Mod.\ Phys.\ Lett.\ A }{\bf #1}}
\def\JPG#1,{{\jnfont J.\ Phys.\ G }{\bf #1},}
\def\CTP#1,{{\jnfont Commun.\ Theor.\ Phys.\ }{\bf #1},}
\def\JHEP#1 {{\jnfont JHEP \ }{\bf #1\ }}
\def\NPPS#1,{{\jnfont Nucl.\ Phys.\ Proc.\ Suppl.\ }{\bf #1},}
\def\btt#1 {{tt$\backslash$#1}}

\newcommand{\etal}{{\it et al.}}

\begin{document}


\title{Associated production of Z boson and a pair of new quarks at the LHC}

\author{YANG Shuo}
\affiliation{Key Laboratory of Frontiers in Theoretical Physics,
Institute of Theoretical Physics, Chinese Academy of Sciences, P.O.
Box 2735, Beijing 100190, China }
\email{shuoyang@itp.ac.cn}

\begin{abstract}

The associated production of $Z$ boson and a pair of new quarks at
the Large Hadron Collider (LHC) is studied. The cross sections for
both sequential fermions and vector-like fermions are presented. It
is found that for sequential fermions
 the cross sections can reach $1\sim 10^2$ fb for heavy quark mass $m_Q$ from 1000 GeV to 200 GeV.
 For vector-like quarks, the cross sections are suppressed by mixing parameter
$\sin\theta_L$. Focusing on process $pp \rightarrow b'b'$, we
investigate the possibility of detecting the $6l+2j$ signal. For a
$b'$ with light mass and a large branching ratio of $b'\rightarrow
bZ$, it is found that only several signal events ( parton level )
can be produced with 1000 fb$^{-1}$ integrated luminosity. Although
the signal events are rare, all the final states are produced
centrally and multi lepton final states are clear at hadron
collider, which could be easily detected.

\end{abstract}


\maketitle

\section{Introduction}

One of the main goals of LHC is searching for new physics beyond the
Standard Model (SM). Many new physics models introduce new fermions
to solve some problem or address some open question of SM. In some
models, new fermions play an important role in electroweak symmetry
breaking or CP violation, and their characters may be different from
the presently known fermions. In little Higgs model \cite{LH}, a
pair of vector-like quarks are introduced in order to cancel the
Higgs one-loop quadratic divergence caused by top quark. In another
model \cite{Liu}, for understanding the Higgs, an extra vector-like
generation of matter is introduced within the framework of
supersymmetry. A heavy fourth generation might help in bringing the
gauge couplings close to a unification point at order of $10^{16}$
GeV \cite{G4GNU}. Discovery of such new fermions would revolutionize
our understanding of electroweak symmetry breaking and some other
basic problems. In this paper, we study the associated production of
$Z$ boson and a pair of new heavy quarks. This process offer the
possibilities to probe the electroweak couplings of new quarks.

This paper is organized as follows. In Sec. II, features of new
quarks and phenomenological constraints on new quarks are briefly
reviewed. In Sec. III, the associated production of $Z$ boson and a
pair of quarks for both sequential quarks and vector-like quarks are
investigated. Finally, we give our conclusions in Sec. IV.


\section{New quarks and their phenomenological constraints}
In addition to three known generations of fermions, new fermions are
introduced in various new physics models. They can
 be chiral or vector-like. In this section, we will
start with a brief description of these two types of new quarks and
then discuss the phenomenological bounds.

One can make a replica of a SM family to get the simplest fourth
generation which is the so-called sequential fermions
\cite{newreport}. The couplings of $Z$ to new sequential quarks
 are the same as those for known quarks. The new
quarks can also be vector-like  \cite{review}, where the left-handed
and right-handed chiral components transform the same under
electroweak gauge group $SU(2)_L\otimes U(1)_Y$. Both vector-like
doublet quarks and vector-like singlet quarks are candidates. For
simplicity and illustrative purpose, we only discuss SM with
vector-like singlet extension in this paper. Introducing a
vector-like pair $\tilde{b}$, $\tilde{b^c}$ with quantum numbers
$(3,1,-2/3)$ and $(\bar{3},1,2/3)$ under $SU(3)_c \times SU(2)_L
\times U(1)_Y$, the mass terms of $b$ quark sector can be presented
as
\begin{align}
{\mathcal L} \supset y_{33}Q_3 \tau_2 \Phi^* b_3^c+ f \tilde{b}
\tilde{b^c} + y_{34}Q_3 \tau_2 \Phi^ * \tilde{b^c} +\text{h.c.},
\end{align}
where $y_{33}$ and $y_{34}$ denote Yukawa couplings, $\Phi$ is the
Higgs doublet, $Q_3=\begin{pmatrix} t_3 \\ b_3\end{pmatrix}^T$ and
$b_3^c$ are the third generation quark doublet and singlet,
respectively. After diagonalizing the mass terms, one get the
physical third generation quark $b$ and the new exotic heavy quark
$b'_v$.

\bea
 b &=& \cos\theta_L b_3-\sin\theta_L \tilde{b}\, \qquad
       b^c=\cos\theta_R b_3^c-\sin\theta_R \tilde{b}^c\, , \\ \nonumber
 b'_{v}&=& \sin\theta_L b_3+ \cos\theta_L \tilde{b} \, \qquad
          b_v^{'c}=\sin\theta_R b_3^c+ \cos\theta_R \tilde{b}^c\,.
\eea Vector-like up-type quark extension is similar
 to this case. Replacing the weak eigenstates by the physical states, one can
get the electroweak interactions of new quarks in vector-like
singlet extended model. We summarize the $Z$-new quark-new quark
interactions vertex for sequential quarks and exotic quarks in
vector-like extended model in Table I. For convenience in
discussion, we denote down-type new quark with the same quantum
number as bottom quark by $b'$, sequential $b'$ quark by $b'_4$ and
exotic $b'$ with large admixture of vector-like quark by  $b'_v$ .
The similar notations are taken for up-type new quark $t'$.

\begin{table}[h]
\begin{tabular}{|c|c|c|}
\hline \hline &sequential fermion & vector-like singlet model
\\ \hline
\hline
$Z \bar{t'} t'$ & $\displaystyle
-\frac{g}{2\cos_W}\gamma_{\mu} (-\frac{4}{3} \sin^2\theta_W+P_L)$ &
$\displaystyle -\frac{g}{2\cos_W}\gamma_{\mu} (-\frac{4}{3}
\sin^2\theta_W+P_Ls_L^2)$  \\
$Z \bar{b'}b'$ & $\displaystyle -\frac{g}{2\cos_W}\gamma_{\mu}
(\frac{2}{3} \sin^2\theta_W-P_L)$  & $\displaystyle -\frac{g}{2\cos_W}\gamma_{\mu}
(\frac{2}{3}\sin^2\theta_W-P_Ls_L^2)$  \\
\hline \hline
\end{tabular}
\caption{$Z$-new quark-new quark interactions for sequential quarks
and exotic quarks. $P_{L,R}=\displaystyle(1\mp\gamma_5)/2$,
$s_L=\sin\theta_L$, $c_L=\cos\theta_L$.}
\end{table}

Now, we turn to the phenomenological constraints on the masses and
mixings of new quarks. The most stringent constraints on sequential
fermions are from "oblique parameters" S, T and U \cite{PDG}. These
constraints can be relaxed by allowing T to vary or fourth
generation masses are not degenerate \cite{constraints,4thandH}.
Recently, ref. \cite{4thandH} has identified a region of particle
masses and mixings for new sequential fermions which are in
agreement with all experimental constraints and has minimal
contributions to oblique parameters. Vector-like fermions do not
contribute to "oblique parameters" in the leading order, and thus
these parameters do not constrain their masses.

 As for direct search limits for new quarks,
the strongest bound on $t'$ is $t'>256$ GeV \cite{CDFquark}, which
comes from CDF by searching for $t'\bar{t'}$ with decay $t'
\rightarrow q+W$. Assuming the branching ratio $BR(b' \rightarrow
bZ)=1$, CDF obtains the bound $m_{b'}>268$ GeV \cite{CDFbprime}.
Note that if the decay branching ratio or lifetime of new quarks are
affected by mixing parameters and mass difference between $t'$ and
$b'$, the mass bounds can be softened. Detailed analysis of
experimental constraints on masses of fourth generation quarks can
be found in ref. \cite{expcons}. In this paper, we will consider a
wide mass parameter space that the masses of new quarks are from 200
GeV to 1000 GeV.

CKM unitary and flavor physics also suppressed the mixing angles
between the extra quarks and the ordinary three generation which
suggest the mixing should be small. As for SM with vector-like
fermions extension, GIM mechanism break down and therefore the most
important consequences are the flavor changing neutral current (
FCNC ) and flavor diagonal neutral currents ( FDNC )
\cite{bargersinglet}. $R_b$ requires the mixing parameter
$\sin\theta_L$ between bottom quark and the new down-type quark
should be small, roughly, $\sin\theta_L < 0.1$. For the mixing
parameter between top and the new $t'$ quark, the constrains are
weaker.


\section{Associated production of Z boson and a pair of new quarks}

\begin{figure}[htbp]
 \includegraphics[width=4in]{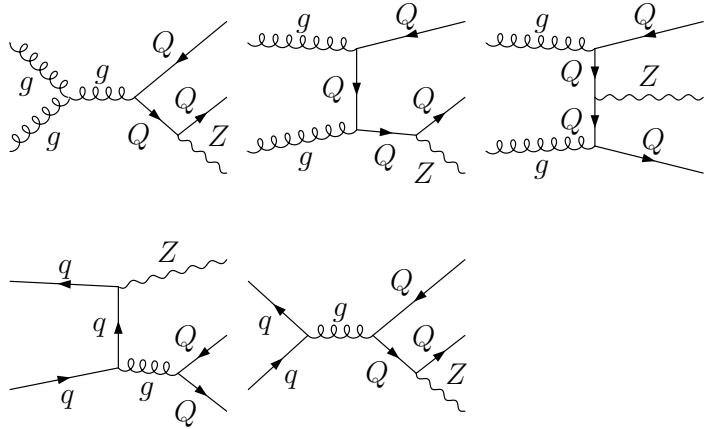}
\caption{Respective Feynman diagrams for associated production of Z
boson and a pair of new quarks. The Feynman diagrams are drawn using
FeynArts \cite{feynarts} .}
\end{figure}

As shown in Fig.1, $Z$ boson can be produced in association with a
pair of new heavy quarks ($pp \rightarrow ZQQ$) at the LHC
\footnote{There also exits the associated production of $Z$ boson
and a pair of new leptons ($pp \rightarrow ZLL$). It originates form
lepton pair production via Drell-Yan mechanism \cite{dy} and
gluon-gluon fusion mechanism \cite{ggLL} with the $Z$ boson emitted
from the lepton lines. However, the cross section for process $pp
\rightarrow ZLL$ is too small.}, which provide a method to detect
the electroweak couplings of new quarks. Probing electroweak top
quark couplings via $t\bar{t}\gamma$ and $t\bar{t}Z$ production at
hadron colliders has been studied \cite{ttz}. In this paper, we
calculate the cross section for process $pp \rightarrow ZQQ$ at the
LHC with $\sqrt{s}=14$ TeV where both chiral quarks ( $t'_4$, $b'_4$
) and vector-like quarks ( $t'_v$, $b'_v$ ) are considered. The
calculations in this paper were performed with MADGRAPH/MADEVENT
\cite{madgraph}. CTEQ6L \cite{cteq6} parton distribution function
with strong coupling constant $\alpha_s(m_Z^2)=0.118$,
renormalization scale and factorization scale $\mu_r=\mu_f=2m_Q$ are
taken. We plot the cross section for the associated production of
$Z$ and $t'$ quarks at the LHC versus mass parameter $m_t'$ in Fig.
2. Both of the cases for chiral quarks and vector-like quarks are
shown for comparison. The cross sections decrease sharply as the
increase of mass parameter $m_Q$. For a sequential $t'$ with mass
from 200 GeV to 1000 GeV, the cross section is variant from $10^2$
fb level to 1 fb level. For a typical $t'$ mass $m_t'=300$ GeV, the
cross sections are 86 fb for sequential $t'$ and 31 fb for
vector-like $t'$, respectively. As for vector-like $t'$, the cross
section is smaller because the coupling is suppressed by $\sin
\theta_L$. Considering phenomenological constraints, we take $\sin
\theta_L\leq 0.1$ and show the cross section for a typical value
$\sin \theta_L = 0.1$ in Fig.2. The results for vector-like fermion
are not very sensitive to the mixing parameter $\sin\theta_L$ in
this region. This is because the cross section is relevant to
$g_v^2+g_a^2$ where $g_v$ and $g_a$ are vector current coupling and
axial current coupling, respectively. And the variation of
$\sin\theta_L$ in the assumed parameter region doesn't affect
$g_v^2+g_a^2$ apparently. So we don't show the cross section for
different value of $\sin\theta_L$ in Fig. 2.

We also plot the cross section for the associated production of $Z$
boson and $b'$ quarks in Fig. 3. The shape of the curves is similar
to those shown in Fig 2. Roughly, the cross section for $ZQQ$
production is relevant to $g_v^2+g_a^2$. For the same mass, the
cross section for associated production of $Z$ boson and $b'$ quarks
is larger than that for $t'$ quarks.

Now we further consider the signature of $ZQQ$ production at the
LHC. The process $pp\rightarrow Zt'\bar{t'}$ followed by
$t'\rightarrow Wq$ results in $ZWWqq$ states , which suffers large
backgrounds from top pair relevant events and gauge bosons events.
For process $pp\rightarrow Zb'\bar{b'}$ with $b'$ decaying to $Z$
and $b$, if one $Z$ decays to neutrinos and another $Z$ decays to
charge leptons, the signal will buried by large $t\bar{t}$
background or $ttZ$ background. If all the $Z$ bosons decay to
hadrons, the signal also suffers large top pair background and it is
hard to trigger at the LHC. So, we concentrate on the $6l+2j$ signal
\footnote{If using $b$ tagging, the 6l+2b signal are suppressed by
$b$ tagging efficiency. Roughly, one $b$ tagging efficiency is
$\epsilon \sim 50\%-60\%$ at the LHC \cite{TDR}. } for $Zb'b'$
production followed by $b'\rightarrow bZ$ where all the $Z$ bosons
decay to charge leptons ( $e$ or $\mu$ ). Although the branching
ratio (BR) of $Z \rightarrow l^+l^-$ is smaller than the BRs for
neutrino decay mode and hadronic decay mode, the charged leptons can
provide efficient trigger and are easily identified at hadron
collider. The process $ZZZjj$ with all $Z$ bosons decaying
leptonically results in $6l+2j$ final states. However, the cross
section for $ZZZjj$ is too tiny and can be neglected. So there is
nearly no background for the $6l+2j$ signal.

\begin{figure}[htbp]
 \includegraphics[width=3.5in]{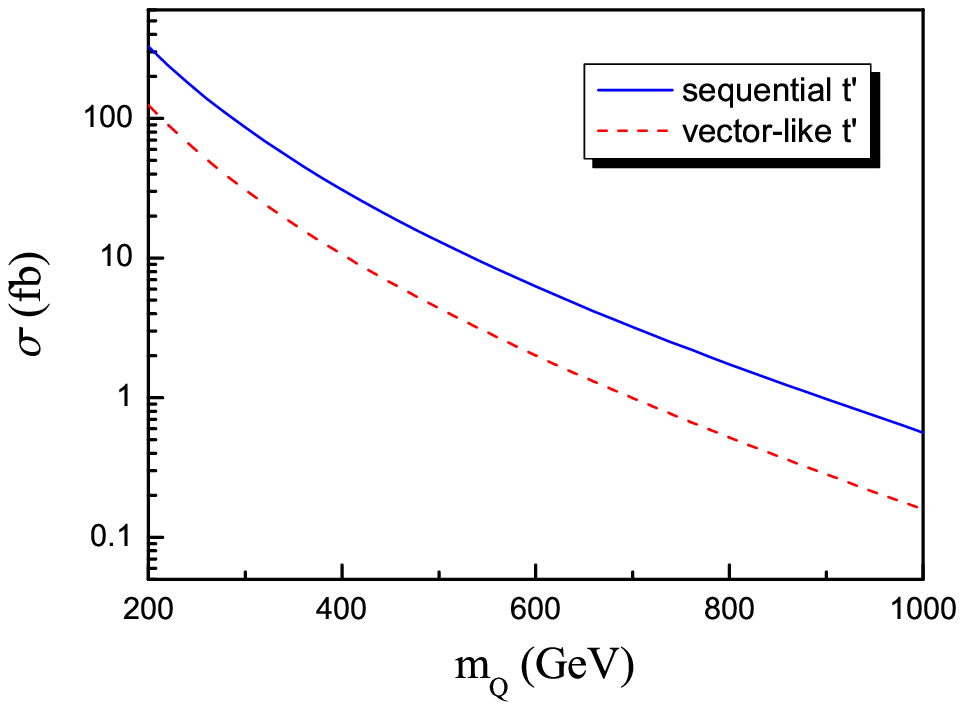}
\caption{Cross section for $t'\bar{t'}Z$ production as a function of
mass parameter $m_Q$. ( solid line for sequential $t'$ and dash line
for vector-like $t'$ )}
\end{figure}

\begin{figure}[htbp]
 \includegraphics[width=3.5in]{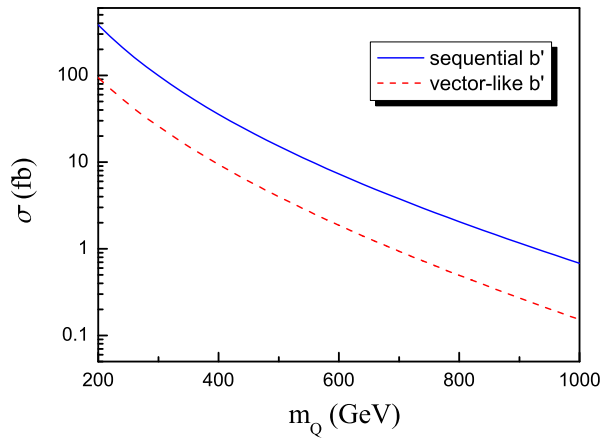}
\caption{Cross section for $b'\bar{b'}Z$ production as a function of
mass parameter $m_Q$. ( solid line for sequential $b'$ and dash line
for vector-like $b'$ )}
\end{figure}

Taking a typical mass of sequential $b'$, $m_Q=250$ GeV, and
assuming the branching ratio $BR(b'\rightarrow bZ) = 50\%$, we
discuss further detecting probability of the signal. We present the
pseudorapidity distribution and transverse momentum distribution of
$b$ quark for signal events in Fig. 5 and Fig 6, respectively. The
kinetic distributions of $\bar{b}$ are the same as those of $b$. As
shown in Fig. 5, the $b$ quarks from $b'$ decay are produced
centrally and tend to reside at high $P_T$ ( peaks at $\sim 90$ GeV
). The leptons from $Z$ bosons decay hold similar features. We also
employ below cuts to simulate detector acceptance. Detector cuts:
\beq P_T(l)> 15 GeV,\ P_T(j)>20 GeV,\ |\eta_j|<5,\ |\eta_l|<5,\
\triangle R(j,j)>0.4,\ \triangle R(l,j)>0.4 .\eeq

\begin{figure}[htbp]
 \includegraphics[width=3.5in]{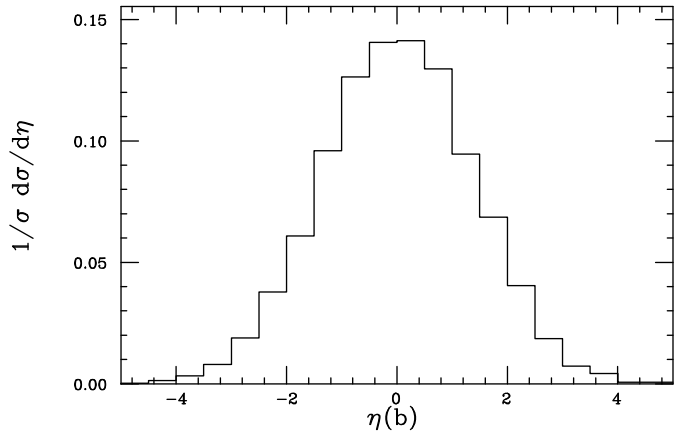}
\caption{ The pseudorapidity distribution of b quark for process
$pp\rightarrow Zb'b'\rightarrow ZbZ\bar{b} Z$ at the LHC.}
\end{figure}

\begin{figure}[htbp]
 \includegraphics[width=3.5in]{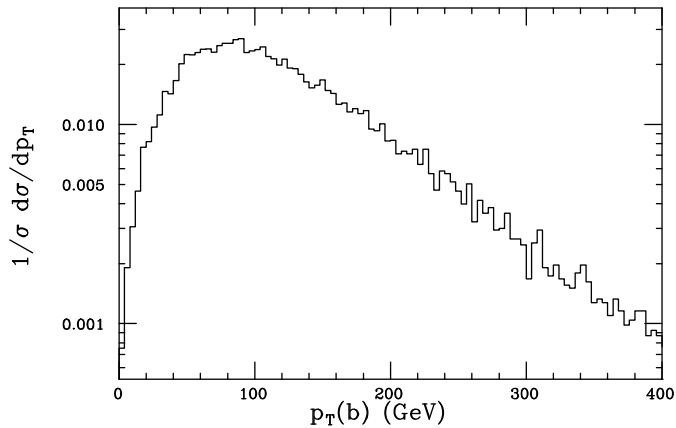}
\caption{The transverse momentum distribution of b quark for process
$pp\rightarrow Zb'b'\rightarrow ZbZ\bar{b} Z$ at the LHC.}
\end{figure}

\begin{table}[h]
\begin{tabular}{|c|c|}
\hline \hline
$$   &  $\sigma$ (fb) \\
\hline
$Z \bar{b'} b'$ & 183.86  \\
$6l+2j$ &  0.0138  \\
Detecor cuts & 0.00844  \\ \hline \hline Events with 1000 fb$^{-1}$& 8\\
 \hline \hline
\end{tabular}
\caption{Cross section (fb) and events number for the signal. }
\end{table}

Because the BR of $Z\rightarrow l^+l^-$ ( $l^-=e,\mu$ ) is small,
the cross section of the signal is suppressed. After including
detector acceptance, the signal events are rare. We give the cross
section and event number for the signal in Table II. It is found
that only 8 signal events ( parton level ) can be produced in the
case we considered with 1000 fb$^{-1}$ integrated luminosity
corresponding to ten years data collection of LHC at high luminosity
$\mathcal{L}= 10^{34}cm^{-2}s^{-1}$. Although the signal events are
rare, all the final states are produced centrally and multi lepton
final states are clear at hadron collider, which could be easily
detected. For larger $m_Q$ and smaller $BR(b'\rightarrow bZ)$, it is
not hopeful to detect the $6l+2j$ signal for process $pp \rightarrow
Zb'b'$. Further research is needed for other signal for process $pp
\rightarrow ZQQ$.


\section{Conclusion}

In this paper, we have considered the associated production of $Z$
boson and a pair of new heavy quarks ( $pp\rightarrow ZQQ$ ) at the
LHC with $\sqrt{s}=14$ TeV. The cross sections for both sequential
fermions ( $t'_4$ and $b'_4$ ) and vector-like fermions ( $t'_v$ and
$b'_v$ ) are presented. For sequential fermions the cross sections
can reach $1 \sim10^2$ fb for heavy quark with mass $m_Q$ from 1000
GeV to 200 GeV. In the case of vector-like quarks, the cross
sections are suppressed by mixing parameter $sin_L$. Focusing on
process $pp \rightarrow b'b'$, we investigate the detecting
possibility for the $6l+2j$ signal. With 1000 fb$^{-1}$ integrated
luminosity corresponding to ten years run of LHC at high luminosity,
only 8 signal events ( parton level ) can be produced in the case we
considered ( $m_Q=250$ GeV and $BR(b'\rightarrow bZ)=50\% $ ).
Although the signal events are rare, all the final states are
produced centrally and multi lepton final states are clear at hadron
collider, which could be easily detected. And only a very narrow
parameter space could be hopeful to detect the $6l+2j$ signal for
process $pp \rightarrow Zb'b'$. For other signal for process $pp
\rightarrow ZQQ$, further research is needed.

\acknowledgments

The author would like to thank Chun Liu for helpful discussions.
This work was supported in part by the National Science Foundation
of China under Grant Nos. 90503002 and 10821504, and by the National
Basic Research Program of China under Grant No. 2010CB833000.

\end{document}